\documentclass[aps,prl,twocolumn,showpacs,showkeys,floatfix,superscriptaddress]{revtex4}

\usepackage{graphicx}
\usepackage{dcolumn}
\usepackage{bm}

\newcommand{\dfrac}[2]{\displaystyle\frac{#1}{#2}}

\newcommand{\condbf}[1]{{#1}}

\begin{document}


\title{Atom-in-jellium equations of state in the high energy density regime}

\date{September 28, 2018, revisions to February 26, 2019 
  -- LLNL-JRNL-752373}

\author{Damian C. Swift}
\email{dswift@llnl.gov}
\affiliation{%
   Lawrence Livermore National Laboratory,
   7000 East Avenue, Livermore, California 94551, USA
}
\author{Thomas Lockard}
\affiliation{%
   Lawrence Livermore National Laboratory,
   7000 East Avenue, Livermore, California 94551, USA
}
\author{Richard G. Kraus}
\affiliation{%
   Lawrence Livermore National Laboratory,
   7000 East Avenue, Livermore, California 94551, USA
}
\author{Lorin X. Benedict}
\affiliation{%
   Lawrence Livermore National Laboratory,
   7000 East Avenue, Livermore, California 94551, USA
}
\author{Philip A. Sterne}
\affiliation{%
   Lawrence Livermore National Laboratory,
   7000 East Avenue, Livermore, California 94551, USA
}
\author{Mandy Bethkenhagen\footnote{%
Present affiliation: Universit\"at Rostock, 18051 Rostock, Germany}}
\affiliation{%
   Lawrence Livermore National Laboratory,
   7000 East Avenue, Livermore, California 94551, USA
}
\author{Sebastien Hamel}
\affiliation{%
   Lawrence Livermore National Laboratory,
   7000 East Avenue, Livermore, California 94551, USA
}
\author{Bard I. Bennett}
\affiliation{%
   Los Alamos National Laboratory,
   PO Box 1663, Los Alamos, New Mexico 87545, USA
}

\begin{abstract}
Recent path-integral Monte Carlo and quantum molecular dynamics simulations
have shown that computationally efficient
average-atom models can predict thermodynamic states in
warm dense matter to within a few percent.
One such atom-in-jellium model has typically been used to predict the 
electron-thermal behavior only, although it was previously developed to
predict the entire equation of state (EOS).
We report completely atom-in-jellium EOS calculations
for Be, Al, Si, Fe, and Mo,
as elements representative of a range of atomic number and low-pressure
electronic structure.
Comparing the more recent method of pseudo-atom molecular dynamics,
atom-in-jellium results were similar: sometimes less accurate, sometimes more.
All these techniques exhibited pronounced effects of electronic shell
structure in the shock Hugoniot which are not captured by
Thomas-Fermi based EOS.
These results demonstrate the value of a hierarchical approach to
EOS construction, using average-atom techniques with shell structure to 
populate a wide-range EOS surface efficiently, complemented by more rigorous 3D
multi-atom calculations to validate and adjust the EOS.
\end{abstract}


\maketitle

\section{Introduction}
Accurate equations of state (EOS) are essential to understand stellar and 
planetary formation and evolution, astrophysical impacts, and engineering
challenges such as the development of inertial confinement fusion energy sources
and the design and interpretation of experiments involving high energy density
(HED)
plasmas such as those using pulsed electrical discharges and laser ablation. 
Experiments to measure the properties of matter in these conditions are
difficult and expensive, and wide-ranging EOS are needed even to design
and interpret such experiments.

Widely-used EOS such as those in the {\sc sesame} and {\sc leos}
libraries \cite{sesame,leos} are usually constructed by combining 
relatively simple semi-empirical models valid over a limited range of states,
such as Thomas-Fermi (TF) or Thomas-Fermi-Dirac (TFD) theory 
\cite{tf,Dirac1928} 
for high compressions and temperatures,
hard-sphere models of the liquid-vapor region \cite{hardsphere},
and measurements of the shock Hugoniot.
Evolving experimental capabilities 
and more rigorous theoretical investigations of localized
regions of the EOS have identified inaccuracies, driving efforts to
construct improved EOS.
However, the most rigorous techniques expected to be applicable for
warm dense matter, 
path-integral Monte-Carlo (PIMC) \cite{pimc} 
and quantum molecular dynamics (QMD) \cite{qmd},
are computationally expensive and not currently practical for the 
direct generation of wide-ranging EOS or for materials of high atomic number 
$Z$.

The most rigorous theoretical techniques simulate the 
kinetic motion of an ensemble of atoms, where the distribution
of the electrons is found with respect to the changing location of the ions
using quantum mechanics \cite{pimc,qmd}.
Despite their rigor, calculations using these techniques are not necessarily
accurate.
Many-body quantum mechanics is based on approximations to address the
problem of representing anticommuting fermion wavefunctions,
the fixed-node approximation \cite{Ceperley1991} in PIMC
and the local density approximation \cite{Hohenberg1964,Kohn1965} to the 
exchange-correlation functional in QMD.
Calculations using either technique are converged to a finite degree 
with respect to numerical
parameters such as the series-sum representation of wavefunctions,
the computation of wavefunctions at a finite set of points in space
(real or reciprocal), and the size of the ensemble of atoms.
The energy of the ensemble is determined from an average over a sufficient
time interval, and the heat capacity can be found from the rate of change
of energy with temperature.
This procedure is computationally expensive, requiring $o(10^{15})$ or more
floating-point operations per state, equivalent to hundreds of CPU-hours
per state for QMD and thousands of CPU-hours for PIMC.
It is typically deemed impractical to perform these simulations for matter
around or below ambient density and above a few tens of electron volts 
using QMD.

Recent PIMC and QMD results have indicated that the simpler approach of
calculating the electron states for a single atom in a spherical cavity
within a uniform charge density of ions and electrons, representing the
surrounding atoms, reproduces their more rigorous EOS
\cite{Benedict2014,Driver2017}.
This atom-in-jellium approach \cite{Liberman1979} 
\condbf{was developed originally to give improved accuracy over 
Thomas-Fermi-based EOS near ambient conditions.}
It was used previously 
to predict the electron-thermal energy of matter at high temperatures and
compressions \cite{sesame,Rozsnyai2001}, as an advance over the approximation of a uniform
electron gas, as in TF and related approaches.

Other techniques are being developed as more advanced compromises between
the accuracy of multi-atom calculations and the efficiency of the jellium
approach, such as orbital-free molecular dynamics \cite{Zerah1992} 
and pseudo-atom molecular dynamics \cite{Starrett2015} (PAMD).
PAMD is based on a higher order representation of electronic states in the
jellium and includes ionic structure self-consistently, 
deducing an effective interatomic potential which can
then be used to perform molecular dynamics (MD) simulations.
PAMD calculations of Be, Al, Si, and Fe produce similar states to
PIMC and QMD \cite{Starrett2016}.
PAMD requires much less computational effort than PIMC or QMD,
but it still amounts to several tens of CPU-hours per state, so the
construction of a wide-ranging tabular EOS is a significant undertaking.

One advantage of these average-atom techniques over QMD is that calculations
of a single atom are fast enough that all electrons can be treated explicitly
under all circumstances. \condbf{For computational efficiency} 
in QMD simulations, the inner electrons are
typically subsumed into a pseudopotential, which 
\condbf{would ideally be fixed and universal over the full range of the EOS.
In practice, for wide-range EOS, the pseudopotential} must be changed or
abandoned at states of very high density or temperature.

Atom-in-jellium theory was previously extended to predict frequencies of
vibration for ions perturbed from equilibrium in the jellium,
and hence the Debye temperature \cite{Liberman1990},
which was assessed as being correct to within $\sim$15\%\ 
for close-packed structures.
The Debye model can be used to predict the ion-thermal free energy,
so this development made it possible in principle to calculate
the complete EOS from atom-in-jellium theory.
However, this does not appear to have been done.

In the work reported here, we made some corrections to the
previous jellium vibrations model,
and calculated EOS to provide a broad comparison
with the more rigorous but more expensive approaches.

\section{Improvements to atom-in-jellium calculations}
The original computer program implementing the atom-in-jellium calculation,
{\sc Inferno} \cite{Liberman1979},
suffered from some numerical problems in convergence and accuracy,
beyond the limitations inherent in the atom-in-jellium model.
For example, {\sc Inferno} experienced convergence problems including failure
to complete calculations at temperatures below 0.1-1\,eV.
To address these problems, a revised program, {\sc Purgatorio}, was written
\cite{Wilson2006}. {\sc Purgatorio} did not however include the ion-thermal 
calculation.
With the help of diagnostics from a variety of {\sc Fortran} compilers available
on different computers, some errors were corrected in {\sc Inferno}, 
including functions
returning incorrect values under some circumstances and machine-dependent
problems arising from the alignment of different types of variables in
common blocks.
The resulting, modified program has been used periodically to calculate
sets of states to help plan HED experiments \cite{Swift1994on}.

{\sc Inferno} is typically used to run a sequence of calculations, and 
its performance on a calculation depends partly on variables set during the
previous calculation. The program was found to perform best when used to 
calculate states along an isochore, starting at the highest temperature of 
interest.

\begin{figure}
\begin{center}\includegraphics[scale=0.72]{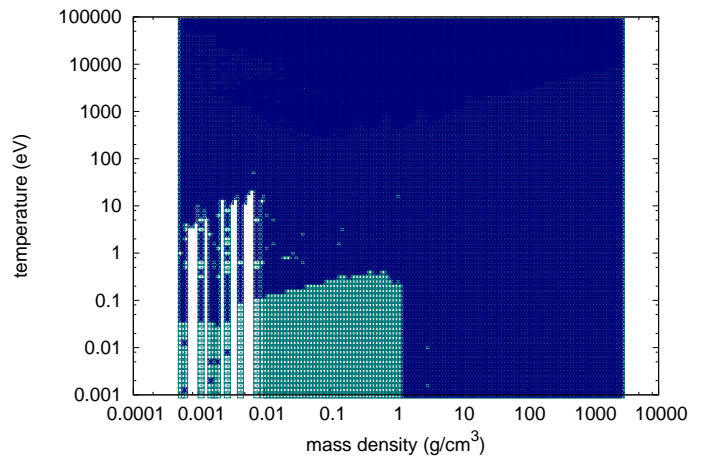}\end{center}
\caption{Atom-in-jellium states calculated for Al.
   Dark blue: electron and ion calculations both completed.
   Light blue: ion calculation failed, indicating non-interacting atoms.}
\label{fig:states}
\end{figure}

Using the modified version of {\sc Inferno}, and calculating down isochores in this way,
atom-in-jellium computations were attempted over a range and density of states
suitable for a general-purpose EOS: mass density $\rho$ from
$10^{-4}$ to $10^3\rho_0$ with 20 points per decade,
and temperature $T$ from $10^{-3}$ to $10^5$\,eV with 10 points per decade.
The electronic wavefunctions were found to be computed reliably down to 10\,K or less for densities
corresponding to condensed matter, and to 100\,K or less for densities
down to 0.1\%\ of the ambient solid.
At lower densities, calculations were completed successfully only for 
temperatures of several eV or more.
Calculations of the ion oscillations tended to fail for densities below 10\%\ of
ambient and temperatures below $\sim$1\,eV, where the electrons were localized
on each atom and an Einstein frequency could not be determined.
The ion-thermal calculation was found to fail or converge inaccurately
for a small fraction of states with no discernible pattern to their
distribution.
(Fig.~\ref{fig:states}.)

The resulting fields 
were post-processed to fill in isolated missing states
and replace obvious numerical glitches,
using polynomial interpolation from surrounding states.
For each state, the Helmholtz free energy $f$ was calculated, and then 
differentiated using a quadratic fit to the three closest states in $\rho$ to
determine the pressure $p(\rho,T)$ in tabular form.
Similarly, quadratic fits in $T$ were differentiated to find the specific
entropy $s$ and hence the specific internal energy $e(\rho,T)$ in tabular form.
These tabulated functions comprise an EOS in {\sc sesame} or {\sc leos} form.

Calculations were performed for a selection of elements desirable for
interpreting HED experiments and for comparison with calculations performed
using other techniques. 
In this paper, we show results for Be, Al, Si, Fe, and Mo.
The first four are interesting to compare with 
recent PAMD results \cite{Starrett2015,Starrett2016} as the closest but more
sophisticated equivalent to the present method.
\condbf{Al, Fe, and Mo are standard materials for which considerable
experimental and theoretical research has been reported.
Al and Mo have been the subject of previous atom-in-jellium studies,
but in a more limited way \cite{Liberman1979}.
The computational cost of calculating a wide-ranging EOS increases much more
quickly with atomic number for PIMC, QMD, and even PAMD, 
than for atom-in-jellium.
Not enough of the EOS has been tabulated to allow the construction of
principal shock Hugoniot over a wide pressure range for Fe or Mo
using these more rigorous techniques. 
We report our construction of wide range EOS for Fe and Mo using the
atom-in-jellium approach, providing predictions which may be tested later when 
the more computationally intensive prescriptions can be applied, or
experimental data may be available.
It is also instructive to compare the predicted evolution of the effects
of ionization of successive electron shells on the shock Hugoniot, with
increasing atomic number.}

\condbf{For consistency with previous atom-in-jellium calculations,}
the exchange-correlation functional was the Hedin-Lundqvist form \cite{Hedin1971}.
\condbf{Calculations using Kohn-Sham \cite{Kohn1965} 
and Perdew-Zunger \cite{Perdew1981} functionals were found to make an
insignificant difference to the EOS in the warm dense matter regime,
compared with the inaccuracy of using the average-atom model instead of
a 3D treatment of the electron distribution.
We would expect gradient-based or hybrid exchange-correlation treatments to make
similarly little difference in this regime, in line with previously-reported results
such as \cite{Sjostrom2016}.}
A single set of solver parameters \cite{Bennett2003}
was used for all calculations.

\condbf{Because the atom-in-jellium model is a simplified representation of
ion and electron distributions in 3D, there is ambiguity in performing some
computations.
The integrals over the continuum electronic states associated with
an atom can be performed over the entire computational domain,
or restricted to the inside of the cavity in the jellium.
Thermodynamic quantities in the model can be defined as the difference
between the calculation for a uniform electron gas with a given
chemical potential, and the calculation for an atom inserted into a cavity
in the uniform electron gas.
The insertion may be performed at constant total volume or at constant
volume of the uniform jellium, the difference being a slightly different
electron density at the boundary.
Calculations were performed with three alternative combinations of these
choices \cite{Liberman1979,Bennett1985}:
\begin{description}
\item[A:] Integrals over continuum states taken over the volume of the cavity,
   cavity inserted into the jellium at constant total volume (compressing
   the jellium as the cavity is added);
\item[B:] Integrals taken over the cavity,
   cavity inserted into the jellium at constant jellium volume (expanding the
   total volume by the volume of the cavity);
\item[T:] Integrals taken over the whole domain,
   cavity inserted into the jellium at constant total volume.
\end{description}
Results from the alternative treatments can be regarded as 
reflecting systematic uncertainties in the atom-in-jellium EOS.}
Typically, the models produce significantly different EOS at low temperatures,
but they converge at temperatures above 1\,eV or so.
Anecdotally, model A has generally been found to be least inaccurate at
low temperatures, but this is not the case for all elements.

The simplest test of a theoretical EOS is how well it reproduces the STP
state.
Multi-atom electronic structure calculations based on variants of the 
local density approximation typically achieve accuracies of $\sim$1\%\ in
lattice parameter, or a few gigapascals in pressure.
Atom-in-jellium results are significantly less accurate,
\condbf{as expected}
(Table~\ref{tab:rho0}).
All calculations used exactly the same solver parameters.
The atom-in-jellium model cannot distinguish solid phases or between
magnetic and non-magnetic structures.

At STP, Al is close-packed and Be is near-close-packed (hexagonal structure
with $c/a$ less than for ideal hexagonal close-packing), and the discrepancy
for both elements is \condbf{relatively small at} a few gigapascals.
Fe and Mo are both body-centered cubic, the former stabilized by magnetic 
ordering; the discrepancy is a few tens of gigapascals, smaller for Mo.
Si is diamond cubic, stabilized by directional covalent-type bonding
which is not captured at all by the atom-in-jellium model,
and has a discrepancy of around 200\,GPa.
\condbf{The discrepancy thus reflects the relative unsuitability of
using a spherical atom-in-jellium treatment for an element exhibiting
a given degree of directional bonding, though the performance for Be
was unexpectedly accurate.}
These discrepancies are a reflection of how far from ambient a material
may have to be in order for the atom-in-jellium calculation to be useful.

\begin{table}
\caption{Pressure calculated at observed STP mass density $\rho_0$ and temperature, for each atom-in-jellium model.}
\label{tab:rho0}
\begin{center}\begin{tabular}{lrrrr}\hline\hline
Element & $\rho_0$ & A & B & T \\
 & (g/cm$^3$) & (GPa) & (GPa) & (GPa) \\ \hline
Be & 1.85 & 6.2 & -1.2 & 11.7 \\
Al & 2.70 & 3.4 & -1.5 & 5.6 \\
Si & 2.33 & -178.5 & -261.0 & -253.0 \\
Fe & 7.86 & -59.1 & -93.1 & -70.2 \\
Mo & 10.28 & -37.5 & -60.6 & -50.1 \\
\hline\hline\end{tabular}\end{center}
\end{table}

\section{Generalized Debye model}
In the ion thermal model developed for use with atom-in-jellium calculations
\cite{Liberman1990},
perturbation theory was used to calculate the Hellmann-Feynman force on the
ion when displaced from the center of the cavity in the jellium.
Given the force constant $k=-\partial f/\partial r$, the Einstein vibration
frequency $\nu_e=\sqrt{k/m_a}$ was determined, where $m_a$ is the atomic mass,
and hence the Einstein temperature $\theta_E=\hbar\nu_e/k_B$.
The Debye temperature $\theta_D$ was inferred from $\theta_E$, either by
equating the ion-thermal energy or the mean square displacement.

\begin{figure}
\begin{center}\includegraphics[scale=0.72]{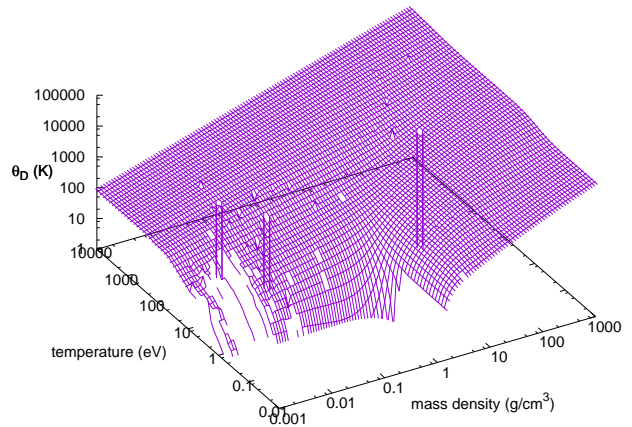}\end{center}
\caption{Debye temperature calculated for Al.}
\label{fig:tdebye}
\end{figure}

The ion displacement calculation can be performed independently at every
$(\rho,T)$ state, and therefore $\theta_E$ and $\theta_D$,
unusually, depend on temperature as well as mass density.
In contrast, in the Debye model of heat capacity as is commonly applied to
condensed matter, $\theta_D$ is assumed to be a function of $\rho$ only.
In practice as calculated using the atom-in-jellium model, 
these characteristic temperatures vary greatly over the wide temperature ranges
applicable to warm dense matter experiments
(Fig.~\ref{fig:tdebye}).
Interpreting the atom-in-jellium results, the electronic states vary as the
atom is heated, and the effect is to increase the restoring force against
displacement of the nucleus from equilibrium.
For all elements studied so far, $\theta_D$ increased much more slowly
than $T$ itself.
The free energy in the Debye model is calculated independently for each $(\rho,T)$ state,
so there is no problem in using $\theta_D$ varying with $T$ as well as $\rho$
in the usual free energy calculation (below) without further modification.

The temperature-dependence of $\theta_D$ in these calculations 
arises solely from thermal excitation of the electrons.
This behavior is distinct from temperature-dependence related to
anharmonicity in the effective interatomic potential, causing
interactions between phonons in the crystal lattice
\cite{anharm}.

Another interesting aspect of the $\theta_D$ calculations is that they
predict an abrupt transition from matter in tension, where the mass density
is lower than at zero pressure but there is still a restoring force on
the displaced nucleus, to instability with respect to perturbations
\condbf{as the electrons localize on the atom 
and effectively cease to interact with neighboring atoms as represented by
the jellium}.
As the temperature is raised \condbf{at constant density}, 
this instability eventually disappears
\condbf{even before ionization takes place, as electrons are promoted to
higher-energy bound states with larger tails at increased radius,
extending more significantly into the jellium}.
The boundary between the two behaviors provides an estimate of the high-density
side of the liquid-vapor region, up to the critical point.
The atom-in-jellium calculation did not include any estimate of cluster 
formation in small groups of atoms, and would not be expected to give
a prediction of the low-density side of the liquid-vapor region.
The displaced-nucleus calculation in effect gives the polarizability of the
jellium, and so is closely related to the van der Waals forces thought
to govern the location of the critical point \cite{Young1971}.
It is difficult to extract a precise prediction of the critical point
from the atom-in-jellium calculations, as they are relatively flat in
$\rho$ and predict a gradual variation from $\theta_D=0$ (a noisy contour) 
to several kelvin
over several thousand kelvin, but the results are broadly consistent with other
estimates, except for the critical temperature of Mo (Table~\ref{tab:crit}).

\begin{table*}
\caption{Critical point.}
\label{tab:crit}
\begin{center}\begin{tabular}{lrrrrl}\hline\hline
Element & \multicolumn{2}{c}{this work} & \multicolumn{2}{c}{literature} & references \\ \cline{2-5}
 & $\rho_c$ & $T_c$ & $\rho_c$ & $T_c$ & \\
 & (g/cm$^3$) & (K) & (g/cm$^3$) & (K) & \\ \hline
Be & 0.20-0.25 & 5000-6500 & 0.25-0.55 & 5300-9200 & \cite{Apfelbaum2012} \\
Al & 0.43-0.70 & 4200-5500 & 0.69 & 7100-8600 & \cite{Young1971} \\
Si & 0.65-1.00 & 7300-7500 &  & 5200 & \cite{ElCat} \\
Fe & 1.00-2.00 & 5900-7800 & 1.33-2.03 & 6750-9340 & \cite{Young1971,Beutl1994} \\
Mo & 0.86-1.31 & 6100-7100 & 1.7-3.7 & 8000-17000 & \cite{Young1971,Minakov2018} \\
\hline\hline\end{tabular}\end{center}
\end{table*}

\begin{figure}
\begin{center}\includegraphics[scale=0.72]{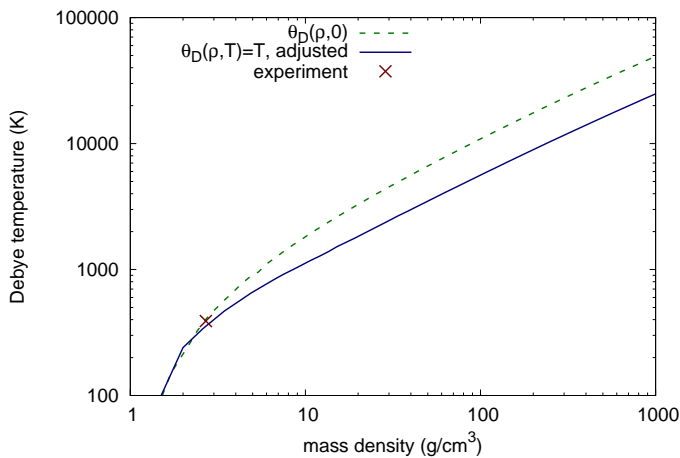}\end{center}
\caption{\condbf{Variation of Debye temperature with
   compression} for Al, 
   \condbf{calculated with atom-in-jellium perturbation theory},
   compared with experimental value
 \cite{Ho1974}.}
\label{fig:tdebt}
\end{figure}

Given $\theta_D(\rho,T)$, the ion-thermal free energy can be calculated from
\begin{equation}
f_i = k_B T\left[3\ln\left(1-e^{-\theta_D/T}\right)+\frac{9\theta_D}{8T}
-D_3(\theta_D/T)\right]
\end{equation}
where $\frac 98 k_B\theta_D$ is the zero-point energy 
and $D_3$ is the Debye integral,
\begin{equation}
D_3(x)\equiv\dfrac 3{x^3}\int_0^x\dfrac{x^3\,dx}{e^x-1}.
\end{equation}
In practice, it was sometimes difficult to correct all the states affected by
numerical noise from $\theta_D(\rho,T)$.
However, the \condbf{precise} value of $\theta_D$ is only important 
\condbf{when the temperature is similar to the Debye temperature.
When $T<<\theta_D$, the ionic heat capacity approaches zero,
and when $T>>\theta_D$, the ionic modes are all saturated.
Thus, in most cases, an adequate representation of the generalized Debye 
heat capacity was found by using}
$\theta_D(\rho):\theta_D(\rho,T)=T$,
i.e. for any $\rho$, the value of $\theta_D$ chosen was that
where it became equal to the temperature.
It was much easier to remove or adjust noisy states from this 
one-dimensional tabulation (Fig.~\ref{fig:tdebt}).

At sufficiently high temperatures, the ions become free and their
specific heat capacity falls from $3 k_B$ to $3k_B/2$ per atom,
where $k_B$ is the Boltzmann constant.
The Debye free energy was modified to account for this freedom 
using a variant of the Cowan model \cite{cowan}.

\section{States at elevated mass density and temperature}
The atom-in-jellium and PAMD methods were originally developed to calculate
states under warm dense matter conditions of compression and 
heating into the plasma regime, which is also most tractable for PIMC and QMD.
All electronic structure methods naturally calculate states at a chosen
mass density and temperature, so comparisons of specific states or loci
where one is held constant, i.e. isotherms and isochores,
are the most direct as they involve the specific, local results from each method.
In contrast, shock Hugoniots involve the initial state as well as
the shock state, which is generally less accurate for the atom-in-jellium 
model,
and isentropes involve either the initial state to establish the
entropy or an integration from the initial state to calculate the
work of compression.
Hugoniot calculations were made using the observed STP mass density,
taking the atom-in-jellium calculation of specific internal energy
(dominated by the binding energy of the inner electrons,
which is likely to be accurate and far greater than the discrepancy in the
outer electrons causing the pressure discrepancy),
but setting the starting pressure to zero if the atom-in-jellium EOS
has a negative value, which is a standard treatment for porous materials.

\subsection{Beryllium}
\begin{figure}
\begin{center}\includegraphics[scale=0.72]{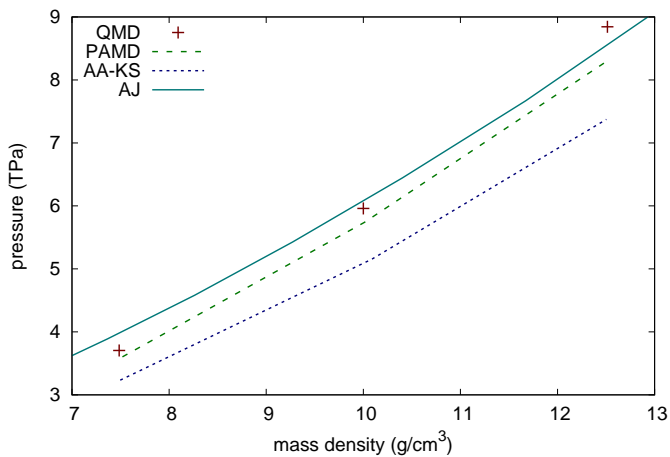}\end{center}
\caption{Comparison of states calculated for Be along the 10\,eV isotherm.
   \condbf{AJ: atom-in-jellium (present work);
   QMD, PAMD, and average-atom Kohn-Sham (AA-KS) \cite{Starrett2016}}.}
\label{fig:beisoth}
\end{figure}

\begin{figure}
\begin{center}\includegraphics[scale=0.72]{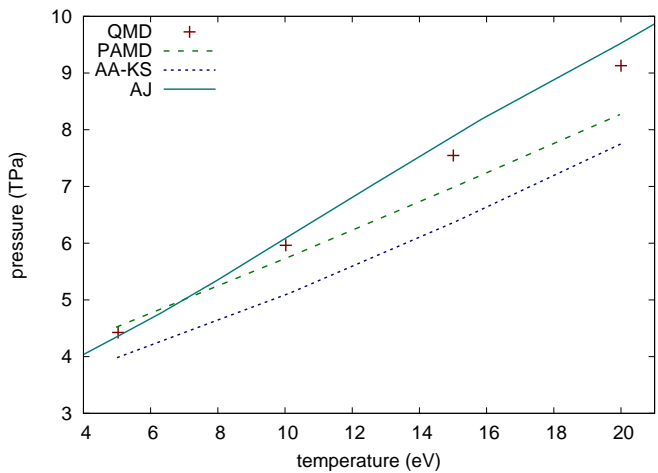}\end{center}
\caption{Comparison of states calculated for Be along the 10\,g/cm$^3$ isochore.
   \condbf{AJ: atom-in-jellium (present work);
   QMD, PAMD, and average-atom Kohn-Sham (AA-KS) \cite{Starrett2016}}.}
\label{fig:beisoch}
\end{figure}

Atom-in-jellium states were extracted from the EOS along
the 10\,g/cm$^3$ isochore and 10\,eV isotherm,
which represent states relevant to ablators used for 
inertial confinement fusion experiments.
The atom-in-jellium results were generally as close or closer to QMD
results, compared with PAMD or previous average-atom Kohn-Sham results
\cite{Starrett2016} (Figs~\ref{fig:beisoth} and \ref{fig:beisoch}).

\begin{figure}
\begin{center}\includegraphics[scale=0.72]{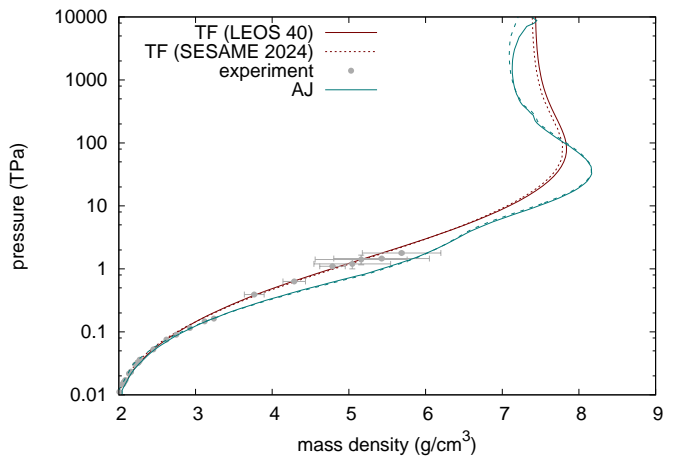}\end{center}
\caption{Shock Hugoniot for Be,
   \condbf{showing comparison between atom-in-jellium (AJ, present work)
   EOS constructed using models A and T (dashed, coincident) and B (solid),
   and TF EOS constructed with slightly different ion-thermal treatments
   \cite{leos,sesame2024},
   with experimental measurements
   \cite{Marsh1980,vanThiel1966,Nellis1997,Ragan1982,Cauble1998}}.}
\label{fig:behugdp}
\end{figure}

The shock Hugoniot from the atom-in-jellium EOS passed closely through
published shock data up to 3.5\,g/cm$^3$ \cite{Marsh1980,vanThiel1966},
lay below the nuclear impedance match data of Nellis et al \cite{Nellis1997},
and passed within the larger error bars of nuclear impedance match and
laser-radiography data at higher pressures \cite{Ragan1982,Cauble1998}.
Previous EOS \cite{leos,sesame2024} were constructed
\condbf{using a straight-line fit to shock speed-particle speed data,
TF theory for the electrons at higher pressure, and different
prescriptions for the ion-thermal energy.
The variation between the TF Hugoniots indicates the sensitivity to
relatively subtle differences in the construction of EOS nominally all based on
the same TF theory.}
The peak compression \condbf{along the Hugoniot, at around 50\,TPa, and not
constrained by existing experimental measurements,} 
was $\sim 6$\%\ higher from the atom-in-jellium calculation 
than from the TF EOS.
\condbf{This is a significant difference, 
and is even larger in terms of pressure, and
should be observable in high pressure shock experiments such as are
becoming possible on high energy pulsed lasers \cite{Kritcher2016}.
At even higher pressures, the different atom-in-jellium models diverged.}
(Fig.~\ref{fig:behugdp}).

\condbf{EOS for Be have been constructed previously using
atom-in-jellium calculations for the electron-thermal excitations only.
One of these EOS included detailed DFT and QMD treatments of the 
solid and liquid phases \cite{Benedict2009};
the cold curve and ion-thermal treatment for the other EOS
\cite{Ding2017} were not reported.
The Hugoniot from our competely atom-in-jellium results was consistent 
with the former EOS, in the liquid and plasma regime,
but differed from the latter.}

\subsection{Aluminum}
\begin{figure}
\begin{center}\includegraphics[scale=0.72]{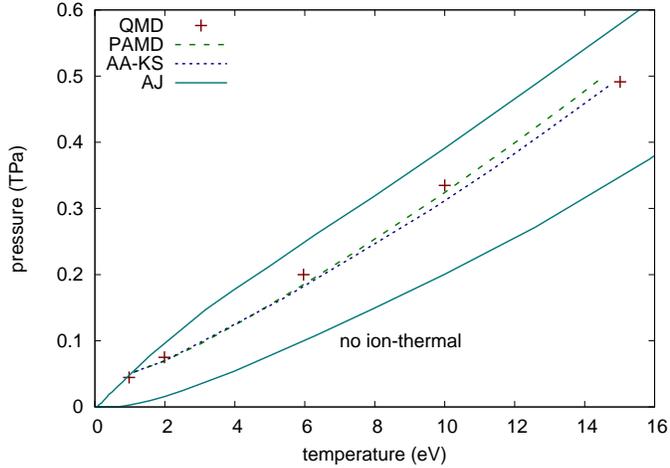}\end{center}
\caption{States calculated for Al along the 2.7\,g/cm$^3$ isochore.
   \condbf{Atom-in-jellium (AJ, present work) is shown with and without
   the ion-thermal contribution, and compared with 
   QMD, PAMD, and average-atom Kohn-Sham (AA-KS) \cite{Starrett2016}.}
}
\label{fig:alisoch}
\end{figure}

Atom-in-jellium states were extracted from the EOS along
the 2.7\,g/cm$^3$ isochore.
The atom-in-jellium results were significantly stiffer than 
previously-reported QMD, PAMD, and average-atom Kohn-Sham results
at these relatively low pressures
\cite{Starrett2016} (Fig.~\ref{fig:alisoch}).
The difference stands out in this comparison because the range is narrower
than for other materials below, and the magnitude of the difference is 
similar at $o(0.1)$\,TPa.
Comparing the total pressure with the contributions from the electrons alone,
the difference could be caused by an overprediction of the ion-thermal
pressure by a few tens of percent.
The difference could be reconciled by a faster decrease in ion-thermal
heat capacity as the kinetic energy of the ions approaches the
binding energy, and thus the attractive potential between the ions becomes
saturated, beyond the Cowan modification to the Debye model.
This will be the subject of a future study.

\begin{figure}
\begin{center}\includegraphics[scale=0.72]{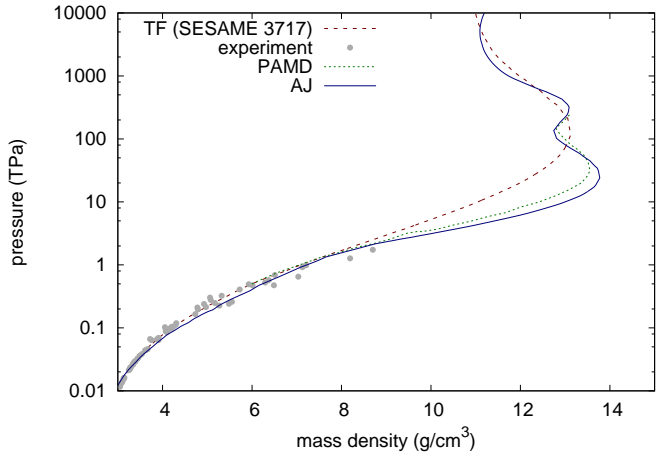}\end{center}
\caption{Shock Hugoniot for Al,
   \condbf{showing comparison between atom-in-jellium EOS (AJ, present work),
   an example TF-based EOS \cite{Barnes1988},
   and PAMD results \cite{Starrett2016},
   with experimental measurements
   \cite{Marsh1980,vanThiel1966,Trunin2001}}.}
\label{fig:alhugdp}
\end{figure}

The shock Hugoniot for Al lay at significantly higher density than observed 
in the solid, but passed within the scatter in the data
\cite{Marsh1980,vanThiel1966,Trunin2001} for Al shocked above
melting.
The behavior closely followed PAMD \condbf{and PIMC} 
calculations \cite{Starrett2016,Driver2018} 
at higher pressures,
exhibiting structure as bound electrons were ionized that departed
significantly from a typical TF-based EOS \cite{Barnes1988}.
(Fig.~\ref{fig:alhugdp}.)

\subsection{Silicon}
\begin{figure}
\begin{center}\includegraphics[scale=0.72]{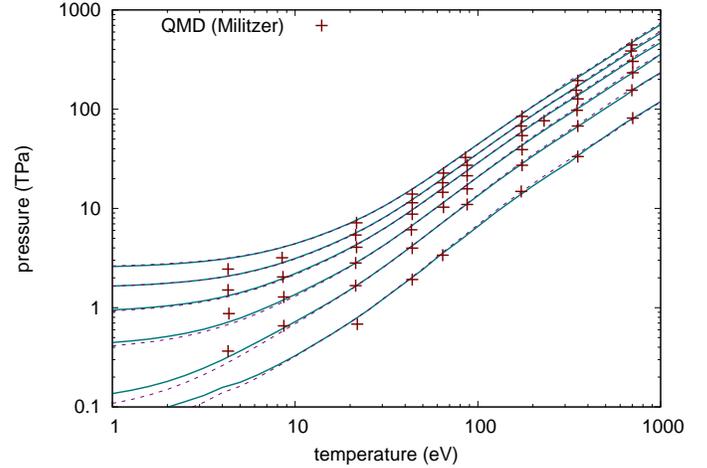}\end{center}
\caption{States in Si along isochores
   at multiples of $\rho_0$ (lowest: $1\times\rho_0$; highest: $6\times\rho_0$)
   \condbf{Atom-in-jellium calculations are shown for models A (dashed) and
      B (solid), compared with previous QMD calculations reported in
      \cite{Starrett2016}.}
}
\label{fig:siisoch}
\end{figure}

Atom-in-jellium states were extracted from the EOS along
isochores from 1 to 6 times $\rho_0$.
Despite the relatively large disagreement with the pressure at STP,
the calculations reproduced previous PIMC and QMD results only slightly less
well than PAMD
\cite{Starrett2016} (Fig.~\ref{fig:siisoch}).
\condbf{This result suggests that a relatively large inaccuracy in
atom-in-jellium predictions in a solid around STP 
does not mean that the EOS will be inaccurate in the warm dense matter regime.
Directional bonds from the outer electrons should disappear
as the atoms become ionized.
Directionality presumably becomes weaker in the liquid; even if present,
it may have a negligible effect 
compared with the typical uncertainties in HED measurements.}

\begin{figure}
\begin{center}\includegraphics[scale=0.72]{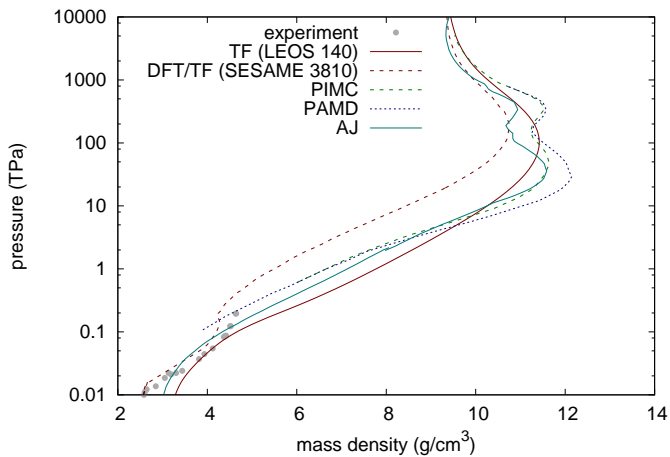}\end{center}
\caption{Shock Hugoniot for Si,
   \condbf{showing comparison between atom-in-jellium EOS (AJ, present work),
   two TF-based EOS \cite{leos,Johnson1997},
   PIMC and PAMD calculations \cite{Starrett2016},
   and experimental measurements
   \cite{Trunin2001,Pavlovskii1968}}.}
\label{fig:sihugdp}
\end{figure}

The shock Hugoniot was very close to results from PIMC \cite{Hu2016},
which had a slightly lower peak compression than
PAMD calculations \cite{Starrett2016}.
All three shell structure techniques predicted features in the Hugoniot as
bound electrons became ionized, differing substantially from TF predictions
\cite{leos,Johnson1997}.
The latter is a combination of multi-ion density functional theory (DFT)
calculations for Si in the diamond phase \cite{Swift1997}
with an empirical piecewise linear representation of the principal Hugoniot
based on that of Ge, to represent low-pressure shock data,
linking to TF at high pressure and temperature.
The resulting Hugoniot exhibits an jump of more than a factor of two
in pressure
\condbf{near the limit in shock data \cite{Trunin2001,Pavlovskii1968} 
around 100\,GPa,
and lies well above the other EOS until above peak compression, 
suggesting that this EOS is probably very inaccurate above 100\,GPa.}
The multi-ion calculations were subsequently extended to include the
$\beta$-Sn phase of Si \cite{Swift2001}.
Calculations have since been performed for other phases,
and a multiphase EOS valid to higher pressures is being constructed.
(Fig.~\ref{fig:sihugdp}.)

\subsection{Iron}
\condbf{QMD studies have been performed for Fe, to study the high-pressure melting curve
for planetary physics \cite{Morard2011,Bouchet2013}, but less has been reported
at higher temperatures relevant to warm dense matter and ionization features along the
shock Hugoniot.}
Here we compare to the sparse QMD and PAMD results previously compared with PAMD calculations
\cite{Starrett2016}
(Table~\ref{tab:festates}).
The atom-in-jellium results were similar to QMD where available and to PAMD,
and the difference lay within the ion-thermal contribution to the EOS,
\condbf{again suggesting that the accuracy could be improved with a more
sophisticated treatment of the reduction in ionic heat capacity from
3 to $\frac 32 k_B$ per atom as the ions become free.}

\begin{table*}
\caption{States in Fe. Previous results from \cite{Starrett2016}.}
\label{tab:festates}
\begin{center}\begin{tabular}{rrrrrrrr}\hline\hline
mass density & temperature & \multicolumn{6}{c}{pressure (TPa)} \\ \cline{3-8}
(g/cm$^3$) & (eV) & QMD & PAMD-KS & TFMD & PAMD-TF & this work & electronic only \\ \hline
18.71 & 5 & 1.61 & 1.560 & - & 2.564 & 1.861 & 1.354 \\
22.50 & 10 & 3.24 & 3.672 & 5.13 & 4.825 & 4.174 & 3.090 \\
34.50 & 100 & - & 66.36 & 68.33 & 67.28 & 69.41 & 59.50 \\
39.65 & 1000 & - & 1456.8 & 1476.5 & 1481.8 & 1497.6 & 1426.6 \\
\hline\hline\end{tabular}\end{center}
\end{table*}

Previous wide-range EOS for Fe include
\condbf{several fitted to shock data where available and merging into TF
theory at high temperature, such as {\sc leos}~260 \cite{leos}.
Iron exhibits solid-solid phase transitions with significant volume change,
which are important for engineering applications involving elevated pressures
and temperatures, such as armor.
For that reason, considerable effort has been devoted to the development
of multi-phase EOS. One well-regarded and wide-range one \cite{Kerley1993}
is a semi-empirical multiphase construction including four solid phases
and the liquid-vapor-plasma region.
This EOS is notable here for using previous atom-in-jellium calculations,
though for the electron-thermal contribution only.}

\begin{figure}
\begin{center}\includegraphics[scale=0.72]{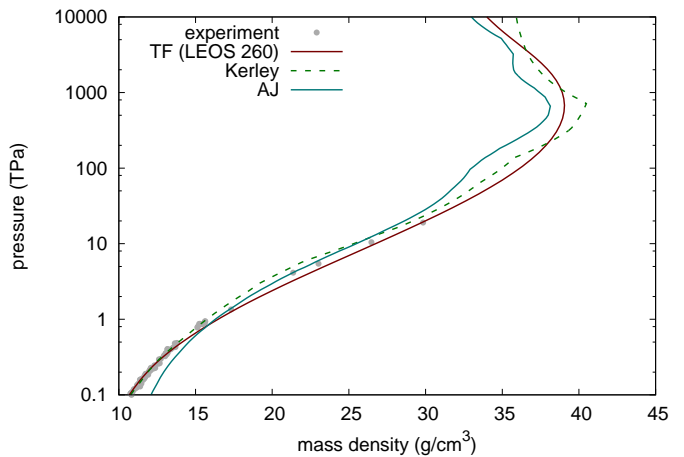}\end{center}
\caption{Shock Hugoniot for Fe,
   \condbf{showing comparison between atom-in-jellium EOS (AJ, present work),
   a semi-empirical EOS incorporating shock data and TF theory \cite{leos},
   Kerley's semi-empirical multiphase EOS which blends into TF theory
      for the cold curve and previous atom-in-jellium
      calculations for the electron-thermal contribution \cite{Kerley1993},
   and experimental measurements
   \cite{Marsh1980,vanThiel1966,Trunin2001}}.}
\label{fig:fehugdp}
\end{figure}

Using the present atom-in-jellium prescription for the whole EOS,
the calculated shock Hugoniot was too dense at pressures below 1\,TPa,
but then followed the (sparse) experimental data
\cite{Marsh1980,vanThiel1966,Trunin2001}
within its scatter.
The shell-structure EOS exhibited very similar modulation in compression
around the peak, though different peak compressions,
\condbf{The TF and multiphase EOS were constructed using cold compression
curves, combined with thermal excitation models for the ions and electrons.
The cold compression curves were algebraic functions,
Birch-Murnaghan in the case of the multiphase EOS \cite{Kerley1993},
with a transition to TF at high density.
It is striking that the TF and multiphase EOS predict similar Hugoniot shapes
around peak compression, the difference being the additional modulation
from the atom-in-jellium shell structure effect in the electron-thermal
contribution to the multiphase EOS.
Average-atom TF-based cold curves ignore the multi-center distribution of the 
nuclear potential,
and the accuracy of
almost all algebraic cold curves is at best unknown in extrapolation to higher
densities than the fitting data, so it seems likely that the present,
completely atom-in-jellium EOS
gives a more accurate prediction of the peak compression.
The difference amounts to a factor or 2-3 in pressure around 100\,TPa,
which should be observable in future HED experiments \cite{Jenei2019}.}
(Fig.~\ref{fig:fehugdp}.)

\subsection{Molybdenum}
Mo is interesting as a relatively high-$Z$ element used as a 
high-pressure standard, for instance in impedance-matching measurements.
It is also notable as being one of a small number of materials for which
EOS were constructed consistently using atom-in-jellium calculations for the
electron-thermal energy, combined with semi-relativistic band structure
calculations of the cold curve, to evaluate nuclear impedance-matching
experiments \cite{Trainor1983}.
Disappointingly, this EOS does not extend to a high enough temperature
for the effects of shell structure to be evident.

\begin{figure}
\begin{center}\includegraphics[scale=0.72]{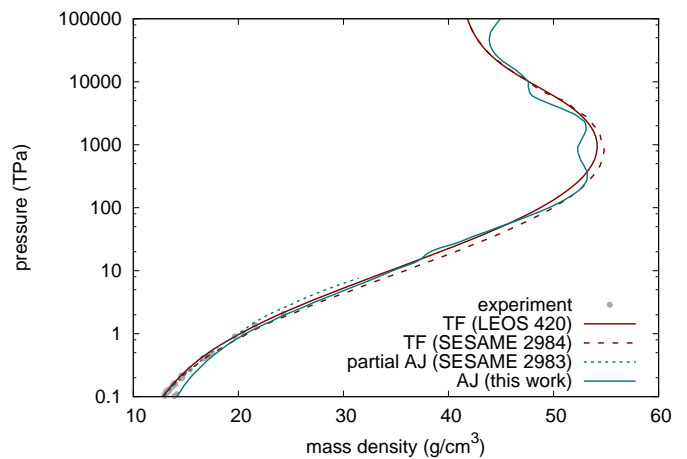}\end{center}
\caption{Shock Hugoniot for Mo,
   \condbf{showing comparison between atom-in-jellium EOS (AJ, present work),
      two semi-empirical EOS using TF theory at high density and temperature
      \cite{Johnson1997a,leos},
      and experimental measurements \cite{Marsh1980,vanThiel1966,Trunin2001}.
      Results from a previous EOS using atom-in-jellium calculations for the
      electron-thermal EOS only \cite{Trainor1983} are also shown,
      demonstrating that the temperature range of this EOS is insufficient
      to show shell structure effects.}
   }
\label{fig:mohugdp}
\end{figure}

\condbf{Wide-ranging semi-empirical EOS have been constructed
using the standard
prescription of an empirical fit to the shock Hugoniot and a transition to
TF theory at high compression and temperatures \cite{Johnson1997a,leos}.}
\condbf{Although constructed using very similar approaches,
the TF-based EOS still differ significantly, particularly in the transition from
the regime constrained by shock data to pressures of several times peak
compression, where the TF contribution dominates equally in both.
Neither PIMC nor PAMD simulations have been reported for Mo,
and QMD simulations have not been reported at states high enough to
explore ionization effects on the shock Hugoniot.
The atom-in-jellium calculation of the complete, wide-ranging EOS
was straightforward, and we include results here as a prediction for
future comparison with more rigorous approaches.}
The present calculations had too high a density at low pressures, but
passed within the scatter in published shock data 
\cite{Marsh1980,vanThiel1966,Trunin2001}
for pressures above 400\,GPa.
At higher pressures, the Hugoniot lay close to the TF EOS but exhibited 
several oscillations as successive electron shells were ionized.
Although relatively modest when plotted over a wide pressure range,
the effects of shell structure still amounted to localized pressure
differences of up to a factor of three in comparison with TF,
\condbf{which should be observable with HED experimental platforms currently under development \cite{Jenei2019}.}
(Fig.~\ref{fig:mohugdp}.)

\section{Discussion}
It may seem surprising that the atom-in-jellium technique published 
in 1979 \cite{Liberman1979} and extended to predict ion-thermal properties
in 1990 \cite{Liberman1990} is still relevant.
However, the standard method of constructing wide-range EOS is still
based on empirical shock wave data in a framework of Mie-Gr\"uneisen and
classical Debye theory (1912) \cite{Debye1912},
with TF theory (1927) \cite{tf} at higher temperatures.
Atom-in-jellium calculations have appeared relatively rarely in EOS libraries
for a variety of reasons, including the higher computational cost than TF,
the unproven predictions of shell structure effects which might be broadened
by the effect of disorder on the ion positions,
and the labor needed to adjust numerical parameters to achieve convergence
or to correct for inadequacies by post-processing individual calculations.
Although the atom-in-jellium states implicitly include the cold contribution
to the EOS, 
average-atom results are usually much less accurate around ambient density than
multi-atom electronic structure calculations, so the atom-in-jellium 
technique has generally been regarded as suitable for 
electron-thermal excitations only, as in the Fe and 
Mo EOS \cite{Kerley1993,Trainor1983}, and even then used rarely.

The updated atom-in-jellium program {\sc Purgatorio} is regarded as
state of the art for calculating the electron-thermal contribution
in wide-range EOS, and the predictions of shell effects on the Hugoniot
are not universally accepted.
{\sc Purgatorio} does not currently include an ion-thermal calculation; 
this contribution is
added to the EOS using a simpler Debye-based model.
Truly general-purpose EOS based on atom-in-jellium calculations 
are not yet widely available
because of the extra complexity involved in combining them with a more
accurate treatment of the solid state, which we have not attempted in the
work reported here.

It is only recently,
with the advent of the more rigorous theoretical techniques,
that the shell structure predictions have been repeated independently, and
over relatively narrow regions of state space.
The LiF EOS whose recent comparisons with PIMC prompted the observation that
shell-structure effects are supported by multi-atom calculations
\cite{Driver2017} is a mixture of two elemental EOS, 
each constructed using {\sc Purgatorio} calculations of the electron-thermal
energy, rather than a direct cross-check without mixing.
Although PIMC is expected to be accurate, it is so computationally 
expensive that its precision has been demonstrated only indirectly,
by comparison with QMD simulations which can themselves be compared with
shock, isothermal compression, or ambient data.
The magnitude of the shell structure effects predicted is only modest
compared with the spacing between states calculated with PIMC,
so the detailed shape may depend partly on the method of interpolation
between simulations.
Clear experimental evidence of the effects of shell structure is still lacking,
though recent developments in converging shock techniques
\cite{Jenei2019,Kritcher2016,Swift2018,Doeppner2018}
mean that direct experimental comparisons may be possible.

A corollary is that EOS construction should proceed via a spectrum of
computational tools, ranging from those capable of spanning a wide range of
state space with as much rigor as is practical, complemented by a hierarchy
of techniques of increasing rigor and computational cost to validate
the EOS or highlight where corrections are needed.
As computational resources and the sophistication of theoretical techniques 
increase, the overall accuracy and rigor of EOS should steadily improve.
Guided by QMD and PIMC in the regimes where they are tractable,
atom-in-jellium calculations are feasible for use now, and appear
necessary to capture expected properties,
for the construction of wide-ranging EOS for elemental plasmas.
For mixed species, atom-in-jellium results must either be mixed
using {\it ad hoc} models, or run with an averaged $Z$, which is physically
dubious because the EOS of each component is highly non-linear in $Z$;
rigorous mixed-species plasma calculations are currently only possible using 
QMD and PIMC.

The calculation of $\theta_D(\rho,T)$ is an interesting generalization of
the Debye model for ion-thermal energy, offering a wider range of validity
than Mie-Gr\"uneisen EOS constructed with $\theta_D(\rho)$.
The related observation that a structure like the vapor dome appears
in atom-in-jellium calculations was unexpected, and provides a different
explanation for the physics behind the liquid-vapor region,
although the EOS calculations themselves do not give a precise calculation
of the critical point.

\condbf{Considering shock Hugoniot curves, 
we showed various comparisons with EOS constructed using the present,
fully atom-in-jellium approach, and other EOS.
The comparisons included semi-empirical EOS
which were constrained by data, usually from shock experiments,
blended into TF or TFD calculations for higher compressions and temperatures.
We also showed a comparison with an example wide-range multi-physics
EOS incorporating
multi-atom DFT and again blended into TF-type calculations outside the range of the 
DFT treatment \cite{sesame}.
Hugoniots vary even from EOS constructed using similar treatments 
by a single researcher on different occasions, 
as well as between different researchers and, even more so,
between different groups using different computer programs.
Hugoniots may vary even in the range of identical sets of
constraining data, as different interpolating functions may be used and 
subsets of the data may be weighted differently when constructing the EOS.
The Hugoniots vary even more markedly between these
approaches and our fully atom-in-jellium calculations.

The deviation between different TF EOS became insignificant for shock
pressures exceeding a few times the level needed to induce peak compression,
because TF-like calculations themselves are relatively standard and equivalent,
and in this regime the TF contribution dominates over differences in
the treatment of the cold compression curve and the ion-thermal excitations.
An exception may be for low-$Z$ materials if the ion-thermal treatment is
particularly crude, when the extra contribution of $\frac 32 k_B$ per atom
to the heat capacity from the potential modes of the ions may still be evident.
The behavior of the EOS depends on the strategy chosen to switch between
different physical approaches in different regimes of state space,
which is not constrained well on physical grounds, and so is particularly
prone to variations between different attempts at constructing a wide-range EOS
for a given substance.
These issues motivate the development of more rigorous theoretical
techniques, but also of computationally efficient techniques enabling
state space to be explored more widely and densely than with the most
rigorous technique available at any juncture.

The atom-in-jellium model was developed precisely to extend the validity of
TFD techniques toward ambient conditions,
by capturing the physics of compressed and heated atoms more accurately
\cite{Liberman1979}.
Atom-in-jellium calculations should not be expected to be as accurate as
PIMC, QMD, or PAMD calculations, 
but our results show that they are likely to be 
adequate for matter in the fluid and dense plasma regime,
though not sufficiently accurate for solid elements near zero pressure.
The more sophisticated techniques have better physical fidelity in
their treatment of ionic motion as the ions transition from being bound,
with fully populated vibrational modes and a heat capacity of $3 k_B$ per atom,
to an unbound state with a heat capacity from the kinetic modes only, and so
falling to $\frac 32 k_B$ per atom.
This is, however, an area with potential for improvements to be made
to atom-in-jellium calculations
\cite{Swift2019}.}

\section{Conclusions}
Equations of state were constructed for five example elements,
using atom-in-jellium calculations
for the ion-thermal as well as electronic free energy.
\condbf{The elements chosen are all solids at STP,
but spanned low to mid $Z$, and a range of types of electronic structure.}
The calculations were efficient enough to allow a wide-ranging EOS to be
produced in a few CPU-hours, and covering compressions and temperatures
typical of general-purpose libraries such as {\sc sesame} and {\sc leos}.
Post-processing is needed as a palliative for numerical noise and 
failed calculations in cool, expanded states.

The calculated states exhibit localization of the electrons at low
temperatures in expansion,
and suggest an atomistic interpretation of the critical
point and boundary of the vapor region as the locus where atoms'
respective electrons start to interact.

The atom-in-jellium EOS were generally inaccurate for states around STP,
particularly for non-close-packed and covalently-bonded structures,
\condbf{but became much more accurate at relatively modest temperatures
or compressions, generally on heating beyond melt.}
For states representative of warm dense matter, the atom-in-jellium
calculations agreed well with PAMD calculations, and gave a similar
agreement with PIMC calculations.
\condbf{PIMC, QMD, and PAMD calculations were more sparse or absent for the 
higher-$Z$ elements, but we made fully atom-in-jellium predictions
of the principal shock Hugoniots for future comparison.}

\condbf{Semi-empirical or multi-physics EOS constrained by shock data or
multi-atom DFT at low temperatures, and blended into TF-type
calculations at higher temperatures, were shown to vary significantly
at shock pressures from the top of the constraining data to pressures
a few times higher than that corresponding to peak compression.
Fully atom-in-jellium calculations are likely to be more accurate in this
regime as they capture more of the physics of the compressed, heated atoms,
though PIMC, QMD, and PAMD calculations are likely to be more reliably
accurate as the ion-thermal energy transitions from fully-populated 
vibrational modes to unbound motion.
There is scope for futher improvement of the treatment of this regime
in atom-in-jellium calculations.}

Shock Hugoniots reproduced the locus of experimental data for pressures
above a few hundred gigapascals, and (like PAMD and PIMC) exhibited
structure as bound electrons were excited.
These structures gave Hugoniots that were significantly different 
from TF-based EOS, though at higher $Z$ the shell structure techniques tended
toward the TF locus though still with significant deviations.
Like PAMD, the atom-in-jellium calculations gave a slightly different
peak Hugoniot compression.
\condbf{The deviations from EOS based on TF theory amount to tens to
hundreds of percent in pressure, and several to $\sim$20\%\ in mass density,
for shock pressures of a few to $\sim$100\,TPa.
This differences should be observable using experimental techniques under
development on HED facilities.}

\section*{Acknowledgments}
Burkhard Militzer (University of California, Berkeley)
kindly provided numerical values from PIMC calculations for comparison.
\condbf{Charles Starrett and Carl Greeff (Los Alamos National Laboratory)
gave extensive input and comments on the manuscript.}

This work was performed under the auspices of
the U.S. Department of Energy under contract DE-AC52-07NA27344.

\end{document}